# Smoothing and extrapolating shifted-contour auxiliary-field Monte Carlo signals using discrete Laguerre functions

**Shlomit Jacobi and Roi Baer**♣

*Fritz Haber Center for Molecular Dynamics, Institute of Chemistry, the Hebrew University of Jerusalem, Jerusalem, 91904, Israel.*

We develop a new smoothing/extrapolating method, based on a discrete Laguerre functions, for systematically analyzing the stochastic signal of shifted-contour auxiliary-field Monte Carlo. We study the statistical errors and extrapolation errors using full configuration-interaction energies for the doubly stretched water molecule. The only parameter at the user's discretion is the order $N$ of the fit. We show that low $N$ emphasizes stability while higher $N$ enable improved extrapolation, at the cost of increased statistical errors. Typically, one should use low order for signals based on a small number of iterations while higher order is efficacious for signals based on large number of iterations. We provide a heuristic algorithm for determining the order to be used and show its utility.

The auxiliary field Monte Carlo (AFMC) [1] is a stochastic numerical method based on the Hubbard-Stratonovich transformation,[2, 3] enabling, among other things, a statistical estimate $E_S(\beta)$ of the following $\beta$-dependent function:

$$E(\beta)=\frac{\langle\Phi|e^{-\beta H}H|\Phi\rangle}{\langle\Phi|e^{-\beta H}|\Phi\rangle}, \qquad (1)$$

which in the limit $\beta\to\infty$ converges to the ground-state energy $E_{gs}$ of a (fermionic/bosonic) many-body Hamiltonian $H$. In this equation $\Phi$ is an approximation to the ground-state wave function. The $\beta$-dependent random signal $E_S(\beta)$ is distributed normally with mean $E(\beta)$ and variance $\sigma^2(\beta)/I$, where $\sigma(\beta)^2$ is the $\beta$-dependent variance of the population sampled by the AFMC process and $I$ is the number of iterations. AFMC has been used in numerous applications for nuclear [4] and molecular electronic structures.[5-7]

A critical addition to AFMC, preserving its formally exact nature is the imaginary shift[8] reducing $\sigma^2(\beta)$ considerably by analytically eliminating the largest (lowest order) noise term in the underlying stochastic process.[9] The method, shifted-contour AFMC (SCAFMC), has been used to calculate ground-state and excited state energies, as well as other properties of molecules.[8-15]

AFMC studies found that the naïve limit $\lim_{\substack{\beta\to\infty\\ I\to\infty}} E_S(\beta,I)=E_{gs}$ cannot be approached in a controlled manner (except in special cases) because $\sigma^2(\beta)$ diverges as $\beta\to\infty$.[15, 16] This is also true of SCAFMC. An estimate of how large $\beta$ should be for a given system depends on the closeness of $\Phi$ to the ground state. Typically, if $\Delta E$, the first excitation gap then one requires $\beta\Delta E\gg 1$. Systems with significant non-dynamical correlation are characterized by relatively small $\Delta E$, requiring large $\beta$, hitting against a hard-wall because of the exponential rise in $\sigma(\beta)^2$ . For applications of AFMC to electronic structure problems, this forms a serious hurdle, leading to the development of various approximations that stabilize AFMC.[16-22] Alternatively one can use multi-reference approaches, which take excited states explicitly into account, thus reducing the required value of $\beta$.[10, 15] In order to help coping with this problem, methods of smoothing the noisy SCAFMC signal $E_S(\beta,I)$ and *extrapolating* to the infinite $\beta$ limit are extremely helpful.

It is the purpose of this paper to present such a signal processing technique and study its utility in actual applications for estimating the electronic ground state energy of molecular systems. As stated above, the mean of the random variable $E_S(\beta)-E_{gs}$ is equal to $E(\beta)-E_{gs}$ and decays monotonically to zero as $\beta$ grows. Therefore, we may expand $E(\beta)-E_{gs}$ using orthonormal functions which decay to zero asymptotically. A natural choice in this respect would be the Laguerre functions $\phi_n(x)=e^{-x/2}L_n(x)$, where $L_n(x)$ are the Laguerre polynomials. We would then write: $E(\beta)-E_{gs}\approx\sum_{n=0}^{N}c_n\phi_n(\varepsilon\beta)$, where $\varepsilon>0$ is an energy scale parameter. However, the numerically calculated SCAFMC signal is composed of *discrete* values $E_k=E_S(k\Delta\beta,I)$ with $k=0,\dots,K$. Thus our analysis makes use not of the continuous but the "discrete" Laguerre polynomials[23]

$$\Lambda_n(x)=e^{-\lambda n}\sum_{\nu=0}^{n}\left(1-e^{\lambda}\right)^{\nu}\binom{n}{\nu}\binom{x}{\nu} \qquad (2)$$

which are orthogonal with respect to the weight $e^{-\lambda x}$ when summed on all the non-negative integers $x=0,1,2\,\dots$ and where $0\le\lambda$ takes the role of $\Delta\beta\varepsilon$. The "discrete Laguerre functions"

$$\psi_n(x)=\sqrt{e^{\lambda(n-x)}(1-e^{-\lambda})}\Lambda_n(x) \qquad (3)$$

are orthonormal $\sum_{x=0}^{\infty}\psi_n(x)\psi_m(x)=\delta_{nm}$ and form the basis for our fit: $E_k-E_\infty\approx\sum_{n=0}^{N}a_n\psi_n(k)$. Of course, we do not know $E_\infty$, which is our estimate for the full CI energy $E_{FCI}$, hence we can estimate it by minimizing the following function:

$$J[\{a_n\}_{n=0}^{N},E_\infty,\lambda]=\sum_{k=0}^{\infty}\left(\sum_{n=0}^{N}a_n\psi_n(k)-\Delta E_k\right)^2, \qquad (4)$$

where $\Delta E_k=E_k-E_\infty$ for $k\le K$ and zero otherwise. Here, $E_\infty$ is. The minimum of $J$ is obtained when $a_n=\sum_{k=0}^{K}\Delta E_k\psi_n(k)$ and $E_\infty=\frac{\sum_{k=0}^{K}O_k^N E_k}{\sum_{k=0}^{K}O_k^N}$, where $O_k^N=1-\sum_{q=0}^{K}\Delta_{\mathrm{qk}}^{\mathrm{N}}$ and $\Delta_{\mathrm{qk}}^{\mathrm{N}}=$

♣ Email: roi.baer@huji.ac.il

$\sum_{n=0}^{N}\psi_n(q)\psi_n(k)$ . $\lambda$ is chosen so that the $K\times K$ matrix $\frac{\partial\Delta_{kq}^{n}}{\partial\lambda}$ $(k,q=0,\dots K)$ is singular and $\Delta E_k$ is its zero eigenvector: $\sum_{k=0}^{K}\frac{\partial\Delta_{qk}^{N}}{\partial\lambda}\Delta E_k=0$.

Two points are here stressed: (1) For a given discrete signal $E_k$ and a required interpolation order $N$ the procedure outlined above determines $a_n$, $\lambda$ and $E_\infty$ without ambiguity. (2) The variational character of the procedure enables using it with analytical derivative for estimating the Born-Oppenheimer force.[24]

For benchmarking we study the water molecule in the equilibrium configuration and in two doubly stretched OH bonds ($R_{OH}$) configurations: with $R_{OH}$ equal to 1, 1.5 and 2 times the equilibrium bond length $R_{OH}^{(0)}=0.978$Å. The bond angle is fixed at 110$^{o}$.[25] We use the DZV basis-set, freezing the core electrons. Such a system is small enough to enable calculation of the full CI energy, for comparison purposes. In all calculations, $\beta_{\max}=10\mathrm{E_h^{-1}}$ and $\Delta\beta=0.1\mathrm{E_h^{-1}}$, thus every SCAFMC iteration involves $\beta_{max}/\Delta\beta\approx 100$ computed Hamiltonian matrix elements. Although the systems have $C_{2V}$ symmetry, we make no use of spatial or spin symmetry. In Figure 1 we plot several discrete SCAFMC signals together with the best-fitted Laguerre tranients.[26] The transients smooth the statistical noise while providing an extrapolated energy $E_\infty$ estimating $E_{FCI}$. The value of $\lambda$ decreases when non-dynamical correlation grows by the stretch of the bond distances. As discussed in detail below, larger $I$ enables higher order $N$.

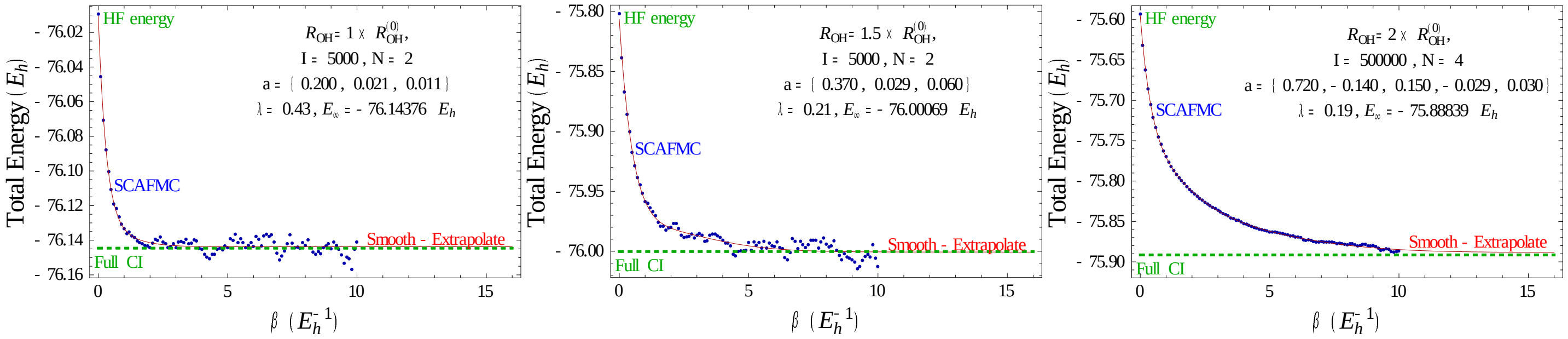

Figure 1: SCAFMC energies (blue dots) as a function of $\beta$ for $I$ iterations, the HF ($\beta=0$) and the full-CI energies (both in green) and the Laguerre-fitted transients for each case (red line) for $H_2O$ with bond lengths ($R_{OH}$), where $R_{OH}^{(0)}\equiv 0.987$Å.

The $\beta\to\infty$ energy estimate as a function of the order $N$ of the fit and the number $I$ of SCAFMC iterations is denoted $E_\infty(N,I)$. It is a random variable distributed with a mean $\mu_\infty$ and a variance $\sigma_\infty^2$. Using a sample of 5 independent SCAFMC signals each of based on $I$ iterations, we can estimate $\mu_\infty$ and $\sigma_\infty$ from the average $\bar{E}_\infty$ and the standard deviation $S_\infty$ of the sample, respectively. We show these estimates in the three panels of Figure 2 for each of the three $H_2O$ configurations. Since for these systems we know the full CI results, we present the extrapolation error $\bar{E}_\infty-E_{FCI}$ instead of $\bar{E}_\infty$ itself. We notice in these results a tradeoff between stability and quality of extrapolation. In general, small $N$ produce the smaller $S_\infty$, stabilizing the statistical error. Looking at the $N=0$ results for $\frac{R_{OH}}{R_{OH}^{(0)}}=1$, we notice that not only the standard deviation is small but the extrapolation error $\bar{E}_\infty-E_{FCI}$ is also small, of the same order. For this system, the $N=0$ fit is already of accuracy of ~0.02eV. The situation is different for the $\frac{R_{OH}}{R_{OH}^{(0)}}=1.5$ system. Here we see that $S_\infty$ is smallest for $N=0$, just as before. For $I=5000$ $S_\infty$ is about 0.1eV and it is comparable to the extrapolation error $\bar{E}_\infty-E_{FCI}$. However, for $I>5000$ $S_\infty$ is small for $N=0$ but the extrapolation error $\bar{E}_\infty-E_{FCI}$ does not decrease. This shows that the fit of order $N=0$ is not flexible enough to improve the extrapolation. Thus, if one produces more SCAFMC iterations, one must simultaneously increasing the flexibility of the fitting function by increasing $N$. Using $I=10^5$ and $N=2$ gives statistical error $S_\infty$ and extrapolation error on the order of 0.01eV. A similar behavior is seen for the toughest system, $\frac{R_{OH}}{R_{OH}^{(0)}}=2$, only now the errors are bigger. In this case, one cannot get better than errors of 0.3eV without going to the N=4 or 6 extrapolation using $I=10^5$ iterations. The combined statistical and extrapolation errors are then less than 0.1eV. One can compare to the statistics of the random variable $E_K$, characterized by $\bar{E}_K$, the average and $S_K$ the standard deviation. The statistical error for the $N=0,2$ fits are typically smaller than $S_K$ by a factor of 3-10 while for $N=4,6$ by a factor 1-3. In general, the higher order the fit the closer the statistical error to $S_K$ of the naïve averaging.

We now suggest a heuristic algorithm for choosing the proper value of $N$: (a) Set a reference accuracy $\Sigma_{\rm ref}$ and $I=1000$. (b) Produce 5 SCAFMC signals of $I$ iterations and determine $\sigma_\infty(N,I)$ for each of order $N=0,2,4,\dots$. (c) Select the highest order $N_*$ for which the standard deviation $S(N_*)$ is close to $\Sigma_{ref}$ (if all $S(N)$ are much larger than $\Sigma_{ref}$ then set $I\to 5I$ and repeat from step (b)). (d) Averages the 5 signals to one signal (representing $5I$ iterations) and fit this signal by a $N_*+2$ polynomial. The value of $E_\infty$ you obtain is your estimate of $E_{FCI}$.

Applying this heuristic on our benchmark systems gives results shown in Table 1. The achievable accuracy is often (but not always) smaller than the reference accuracy.

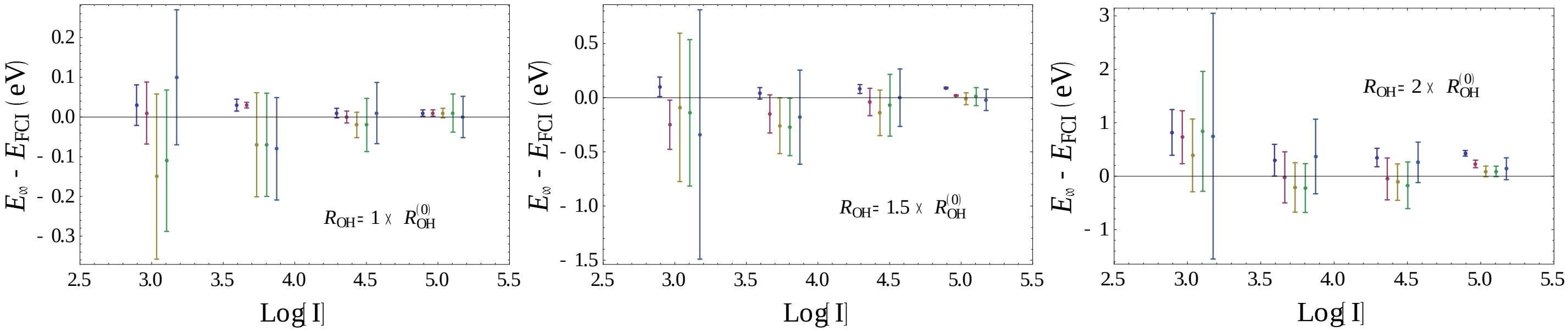

Figure 2: The difference of SCAFMC absolute energy estimates from full CI results and statistical error bars +/I standard deviations at the DZV basis set level (core electrons are frozen) for three $H_2O$ configurations having the $C_{2v}$ symmetry and (equilibrium) bond angle but differing by the OH distances, $R_{OH}$ ($R_{OH}^{(0)} = 0.978$Å). Results are shown for 1000, 5000, 25,000, and 100,000 iterations, and for each number of iterations we give results for 4 interpolating polynomial orders, from left to right: $N = 0,2,4,6$ and the last error bar is obtained from averaging $E_K$, i.e. without a Laguerre fit (we slightly shift horizontally the results for better view).

Summarizing, we have presented a method for smoothing and extrapolating the discrete signals produced by SCAFMC using discrete orthogonal Laguerre functions of order $N$. No essential approximation are being made in the approach and increasingly exact results can be obtained by increasing the number of iterations $I$, increasing $\beta_{max}$ and decreasing $\Delta\beta$ and the fit order $N$. As mentioned above, a very attractive property of the proposed method is its variational character. As will be discussed in a future publication, this allows calculating Born-Oppenheimer forces and response properties by direct sampling of the analytical derivatives of the Hamiltonian. Using the method, we showed that estimates close to full-CI energies (less than 0.1eV) can be obtained with $10^3$ MC iterations for the water in equilibrium, ~$10^4$ MC iterations for mildly stretched bonds and $10^5$ iterations for strongly stretched bonds.

Table 1: Deviance of total energy estimates from full-CI results based on the heuristic algorithm discussed in the text.

| $\frac{R_{OH}}{R_{OH}^{(0)}}$ | Iterations: | 5000 | 25000 | 125000 | 500000 |
|---|---|---|---|---|---|
| 1 | $\Sigma_{ref}$ (eV) | 0.02 | 0.02 | 0.02 | 0.01 |
| | N* | 0 | 2 | 2 | 4 |
| | $E_\infty - E_{FCI}$ (eV) | 0.021 | -0.008 | -0.015 | 0.008 |
| 1.5 | $\Sigma_{ref}$ (eV) | 0.1 | 0.1 | 0.05 | 0.01 |
| | N* | 0 | 0 | 0 | 2 |
| | $E_\infty - E_{FCI}$ (eV) | 0.002 | -0.087 | -0.010 | 0.014 |
| 2 | $\Sigma_{ref}$ (eV) | 0.4 | 0.3 | 0.15 | 0.1 |
| | N* | 0 | 0 | 0 | 2 |
| | $E_\infty - E_{FCI}$ (eV) | 0.80 | -0.01 | 0.09 | 0.10 |

**Acknowledgements:** We gratefully acknowledge the support of this study by the Israel Science Foundation.

1. G. Sugiyama and S. E. Koonin, Ann. Phys. N.Y. **168**, 1 (1986).
2. R. Stratonovich, Dokl. Akad. Nauk. SSSR **115**, 1907 (1957).
3. J. Hubbard, Phys. Rev. Lett. **3** (2), 77 (1959).
4. S. E. Koonin, D. J. Dean and K. Langanke, Phys. Rep. **278** (1), 2-77 (1997).
5. S. Sorella, S. Baroni, R. Car and M. Parrinello, EPL (Europhysics Letters) **8** (7), 663 (1989).
6. P. L. Silvestrelli, S. Baroni and R. Car, Phys. Rev. Lett. **71** (8), 1148-1151 (1993).
7. D. M. Charutz and D. Neuhauser, J. Chem. Phys. **102** (11), 4495-4504 (1995).
8. N. Rom, D. M. Charutz and D. Neuhauser, Chem. Phys. Lett. **270** (3-4), 382-386 (1997).
9. R. Baer, M. Head-Gordon and D. Neuhauser, J. Chem. Phys. **109** (15), 6219-6226 (1998).
10. N. Rom, E. Fattal, A. K. Gupta, E. A. Carter and D. Neuhauser, J. Chem. Phys. **109** (19), 8241-8248 (1998).
11. R. Baer, Chem. Phys. Lett. **324** (1-3), 101-107 (2000).
12. R. Baer, J. Chem. Phys. **113** (2), 473-476 (2000).
13. R. Baer and D. Neuhauser, J. Chem. Phys. **112** (4), 1679-1684 (2000).
14. R. Baer, Chem. Phys. Lett. **343** (5-6), 535-542 (2001).
15. S. Jacobi and R. Baer, J. Chem. Phys. **120** (1), 43-50 (2004).
16. E. Y. Loh, Jr., J. E. Gubernatis, R. T. Scalettar, S. R. White, D. J. Scalapino and R. L. Sugar, Phys. Rev. B **41**, 9301 (1990).
17. E. Y. Loh, Jr. and J. E. Gubernatis, in *Electronic Phase Transitions*, edited by W. Hanke and Y. V. Kopaev (North-Holland, Amsterdam, 1992), Vol. 32.
18. S. B. Fahy and D. R. Hamann, Phys. Rev. Lett. **65** (27), 3437-3440 (1990).
19. S. W. Zhang, J. Carlson and J. E. Gubernatis, Phys. Rev. Lett. **74** (18), 3652-3655 (1995).
20. D. R. Hamann and S. B. Fahy, Phys. Rev. B **47** (4), 1717-1725 (1993).
21. S. W. Zhang, J. Carlson and J. E. Gubernatis, Phys. Rev. B **55** (12), 7464-7477 (1997).
22. Y. Asai, Phys. Rev. B **62** (16), 1610674 (2000).
23. M. J. Gottlieb, Am. J. Math. **60**, 453 (1938).
24. This will be discussed in a future publication.
25. R. A. Phillips, R. J. Buenker, P. J. Bruna and S. D. Peyerimhoff, Chem. Phys. **84** (1), 11-19 (1984).
26. While the Laguerre functions are orthogonal with respect to the integers, they are well-defined for non-integer numbers (using the Gamm function), and thus can be represented as continuous lines in Figure 1.